\documentclass[journal]{IEEEtran}
\IEEEoverridecommandlockouts
\usepackage{cite}
\usepackage{hyperref}
\usepackage{amsmath,amssymb,amsfonts}
\usepackage{amsmath}
\usepackage{amsthm}

\usepackage{graphicx}
\usepackage{textcomp}
\usepackage{xcolor}
\usepackage{graphicx}
\usepackage{float}
\usepackage{subfigure}
\usepackage{amsmath}
\usepackage{amsfonts,amssymb}
\usepackage{mathrsfs}
\usepackage{mathtools}
\usepackage{algorithm}
\usepackage{algorithmicx}
\usepackage{algpseudocode}
\usepackage{bm}
\usepackage{multirow}
\usepackage{array}
\newtheorem{theorem}{Theorem}
\makeatletter
\newcommand{\biggg}{\bBigg@{3}}
\newcommand{\Biggg}{\bBigg@{3.5}}
\makeatother
\usepackage{stfloats}

\bibliography{IEEEabrv}
\def\BibTeX{{\rm B\kern-.05em{\sc i\kern-.025em b}\kern-.08em
    T\kern-.1667em\lower.7ex\hbox{E}\kern-.125emX}}

\begin{document}

\title{Physical Layer Security Over Mixture Gamma Distributed Fading Channels With Discrete Inputs: A Unified and General Analytical Framework\\
}

\author{Chongjun~Ouyang,
        Sheng~Wu,
Chunxiao~Jiang,~\IEEEmembership{Senior Member,~IEEE,}\\
Julian~Cheng,~\IEEEmembership{Senior Member,~IEEE,}
and Hongwen~Yang
\thanks{C. Ouyang, S. Wu, and H. Yang are with the School of Information and Communication Engineering, Beijing University of Posts and Telecommunications, Beijing 100876, China. E-mail: \{DragonAim, thuraya, yanghong\}@bupt.edu.cn.}
\thanks{C. Jiang is with Tsinghua Space Center, Tsinghua University, Beijing 100084, China, and with Beijing National Research Center for Information Science and Technology, Beijing 100084, China. E-mail: jchx@tsinghua.edu.cn.}
\thanks{J. Cheng is with the School of Engineering, The University of British Columbia, Kelowna, BC V1V 1V7, Canada. Email: julian.cheng@ubc.ca.}
}

\maketitle

\begin{abstract}
Physical layer security is investigated over mixture Gamma (MG) distributed fading channels with discrete inputs. By the Gaussian quadrature rules, closed-form expressions are derived to characterize the average secrecy rate (ASR) and secrecy outage probability (SOP), whose accuracy is validated by numerical simulations. To show more properties of the finite-alphabet signaling, we perform an asymptotic analysis on the secrecy metrics in the large limit of the average signal-to-noise ratio (SNR) of the main channel. Leveraging the Mellin transform, we find that the ASR and SOP converge to some constants as the average SNR increases and we derive novel expressions to characterize the rates of convergence. This work establishes a unified and general analytical framework for the secrecy performance achieved by discrete inputs.
\end{abstract}

\begin{IEEEkeywords}
Discrete inputs, mixture Gamma distribution, physical layer security, secrecy performance analysis.
\end{IEEEkeywords}

\section{Introduction}
Due to the broadcast nature of radio propagation, wireless communication networks have been suffering from serious information leakage that arises from eavesdropping attacks. Given this backdrop, physical layer security (PLS) has emerged as an attractive paradigm for secure communication \cite{b1}. In \cite{b2}, Wyner showed that leveraging a pair of properly designed encoder-decoder, PLS approach can achieve perfect secrecy where the eavesdropper cannot obtain any information about the message dedicated to the legitimate receiver. Hence, the issue of PLS has received considerable research attention.

Over the past years, secrecy performance analysis over fading channels has become a hot research focus in area of PLS, and such an analysis can unveil valuable system design insights. Specifically, the secrecy performance measurement metrics including the average secrecy rate (ASR) and secrecy outage probability (SOP) have been analyzed over the simple small-scale fading channels, e.g., the Rayleigh \cite{b3}, Nakagami-$m$ \cite{b4}, $\eta$-$\mu$ \cite{b5}, and $\kappa$-$\mu$ \cite{b6} models, {the large-scale fading channels, e.g., the $M$-distributed model \cite{b21}, the cascaded fading channels, e.g., the double-Rayleigh model \cite{b22}}, and the composite fading channels, e.g., the Generalized-$K$ (${\mathcal K}_G$) \cite{b7} model. Recently, it was found that the mixture Gamma distribution (MGD) serves as a general model to characterize the received signal-to-noise ratio (SNR) of various fading types \cite{b8}, including many classical fading distributions, such as Rayleigh, Nakagami-$m$, $\eta$-$\mu$, $\kappa$-$\mu$, and ${\mathcal K}_G$. The secrecy performance over the MGD fading model was studied in \cite{b9}. Due to the wide generality of the MGD model, the work in \cite{b9} serves as a generalized framework of \cite{b3,b4,b5,b6,b7}.

The derivations in the aforementioned works were based on the assumption that the transmitter leveraged a Gaussian random codebook to encode the secret message \cite{b10}. In a nutshell, these works have laid a solid foundation for understanding the secrecy risks in the wireless physical layer exploiting Gaussian inputs. Yet, in practical systems, the symbols in wiretap codes are taken from a set of discrete finite alphabets, which makes the input signals non-Gaussian in general \cite{b11}. {Thus, it makes more sense to analyze the ASR and SOP of finite input signals. Yet, research on the PLS under discrete inputs focused more on the precoding design \cite{b11} and the problem of secrecy performance analysis has received scant attention.}

To fill this knowledge gap, this work performs both explicit and asymptotic analyses to the ASR and SOP achieved by finite-alphabet signaling. {As previously stated, many classical fading distributions can be characterized by the MGD model. To establish a unified and general analytical framework for secrecy performance evaluation over these mixture Gamma distributed fading channels \cite{b8}, we take the MGD model into consideration. For other fading channels where the instantaneous received SNRs cannot be rewritten in terms of the mixture Gamma distribution such as the $M$-distributed \cite{b21} and double-Rayleigh \cite{b22} fading channels, their secrecy performance will be analyzed in our future works.}

\section{System Model}
In a classical Alice-Bob-Eve wiretap fading channel \cite{b2}, the transmitter (Alice), the legitimate receiver (Bob), and the eavesdropper (Eve) are all single-antenna devices\footnote{We note that the statistics of the singular values in a random matrix with mixture Gamma distributed elements still remain as open issues, which makes it challenging to analyze the secrecy performance of multiple-input multiple-output (MIMO) channels. Thus, this work considers single-input single-output (SISO) channels and the MIMO cases will be considered in our future works.}. The received signals at Bob and Eve can be written as
\begin{equation}
y_i=h_i s+z_i,
\end{equation}
where $z_i\sim{\mathcal{CN}}\left(0,1\right)$ ($i\in\{{{\text B}},{{\text E}}\}$) denote the additive white Gaussian noises, $h_i{\in}{\mathbb C}$ denote the channel gains, and $s$ denotes the transmitted symbol satisfying ${\mathbb E}\left\{\left|s\right|^2\right\}=1$. Assume that the transmitted symbols are taken from an $M$-ary quadrature amplitude modulation (QAM) constellation $\mathcal S$ with equal probabilities. Aided with \cite{b12,b13}, the instantaneous mutual information over the main and eavesdropper's channels can be written as ${{\mathcal I}}_M\left(\gamma_i\right)=\log_2\frac{M}{{\rm e}}-{{\mathcal L}}_M\left(\gamma_i\right)$, where
\begin{equation}\label{EQUATION0}
\begin{split}
{{\mathcal L}}_M\left(\gamma_i\right)=&\frac{1}{M{\pi}}\sum\nolimits_{j=0}^{M-1}\bigg[\int_{\mathbb C}{\rm e}^{-{\left|u-\sqrt{\gamma_i}s_j\right|^2}}\\
&\times\log_2\left(\sum\nolimits_{k=0}^{M-1}{\rm e}^{-{\left|u-\sqrt{\gamma_i}s_k\right|^2}}\right){\rm d}u\bigg].
\end{split}
\end{equation}
Here, $\mathbb C$ represents the complex plane, $s_j\in{\mathcal S}$ denote the $M$-QAM symbols, and $\gamma_i=\left|h_i\right|^2$ denote the instantaneous received SNRs. Generally, ${{\mathcal L}}_M\left(\cdot\right)$ lacks any close-form expressions and can be only calculated by methods of numerical integration \cite{b11}. Yet, for the commonly used square $M$-QAM signals, a closed-form approximation of ${{\mathcal L}}_M\left(\gamma_i\right)$ is available. Particularly, for square $M$-QAM signals, we can obtain \cite{b12}
\begin{equation}\label{EQUATION2}
\begin{split}
{{\mathcal L}}_M\left(\gamma_i\right)=&\frac{2}{\sqrt{M\pi}}\sum\nolimits_{j=0}^{\sqrt{M}-1}\bigg[\int_{-\infty}^{+\infty}{\rm e}^{-{\left(u-\sqrt{\gamma_i}p_j\right)^2}}\\&\times\log_2\left(\sum\nolimits_{k=0}^{\sqrt{M}-1}{\rm e}^{-{\left(u-\sqrt{\gamma_i}p_k\right)^2}}\right){\rm d}u\bigg],
\end{split}
\end{equation}
where $p_j$ denotes the real part of $s_j$. On the basis of the Gauss-Hermite quadrature rule \cite[eq. (25.4.46)]{b14}, we have
\begin{equation}\label{EQUATION1}
\begin{split}
{{\mathcal L}}_M\left(\gamma_i\right)\approx&{\hat{\mathcal L}}_M^{\left(n\right)}\left(\gamma_i\right)=\frac{2}{\sqrt{M\pi}}\sum\nolimits_{j=0}^{\sqrt{M}-1}\sum\nolimits_{l=1}^{n}\omega_l\\
&\times\log_2\left(\sum\nolimits_{k=0}^{\sqrt{M}-1}{{\rm e}^{{-\left(t_l+\sqrt{\gamma_i}p_{jk}\right)^2}}}\right),
\end{split}
\end{equation}
where $p_{jk}=p_j-p_k$; $\left\{\omega_l\right\}$ and $\left\{t_l\right\}$ denote the weight and abscissas factors of the Gauss-Hermite integration. We comment that a larger value of $n$ yields a higher approximation precision. {By numerical simulation, we find that setting $n=20$ can generally ensure ${\epsilon_M^{\left(n\right)}}\left(\gamma\right)\triangleq\left|{{\mathcal L}}_M\left(\gamma\right)-{\hat{\mathcal L}_M^{\left(n\right)}}\left(\gamma\right)\right|={\mathcal O}\left(10^{-5}\right)$.}

Based on \cite{b3}, the instantaneous secrecy rate is given by
\begin{align}
{\mathcal I}_{M}^{\text{s}}&=\max\left\{{\mathcal I}_M\left(\gamma_{\text B}\right)-{\mathcal I}_M\left(\gamma_{\text E}\right),0\right\}\label{EQUATION4}\\
&=\max\left\{{\mathcal L}_M\left(\gamma_{\text E}\right)-{\mathcal L}_M\left(\gamma_{\text B}\right),0\right\}.\label{EQUATION5}
\end{align}
We consider $\gamma_i$ follow the mixture Gamma distribution with probability density functions (PDFs) and cumulative distribution functions (CDFs) respectively given by \cite{b8}
\begin{align}
f_i\left(\gamma_i\right)&=\sum\nolimits_{l=1}^{L_i}\alpha_{i,l}\gamma_i^{\beta_{i,l}-1}{\rm e}^{-\zeta_{i,l}\gamma_i},\quad\gamma_i\geq0,\label{EQUATION6}\\
F_i\left(\gamma_i\right)&=\sum\nolimits_{l=1}^{L_i}\alpha_{i,l}\zeta_{i,l}^{-\beta_{i,l}}\Upsilon\left(\beta_{i,l},\zeta_{i,l}\gamma_i\right),\label{EQUATION7}
\end{align}
where $L_i$, $\alpha_{i,l}$, $\beta_{i,l}$, and $\zeta_{i,l}$ denote the fading parameters satisfying {$\int_{0}^{+\infty}f_i\left(\gamma_i\right){\rm d}\gamma_i=\sum_{l=1}^{L_i}{\alpha_{i,l}\Gamma\left(\beta_{i,l}\right)}{\zeta_{i,l}^{-\beta_{i,l}}}=1$ \cite{b8}; $\Gamma\left(z\right)\triangleq\int_{0}^{\infty}{\rm e}^{-t}t^{z-1}{\rm d}t$ is the Gamma function \cite[eq. (8.310.1)]{b15}}; $\Upsilon\left(a,\rho\right)\triangleq\int_{0}^{\rho}t^{a-1}{\rm e}^{{-t}}{\rm{d}}t$ is the lower incomplete Gamma function \cite[eq. (8.350.1)]{b15}. By \cite{b8}, the MGD serves as a general model to characterize the received SNRs of various types of fading channels and the values of $L_i$, $\alpha_{i,l}$, $\beta_{i,l}$, and $\zeta_{i,l}$ depend on the specific target fading type.

\section{Average Secrecy Rate}
\label{sec3}
\subsection{Explicit Analysis}
According to \cite{b16}, ${\mathcal I}_M\left(\gamma\right)$ is monotone increasing. Consequently, the ASR can be written as
\begin{align}
{\bar{{\mathcal I}}}_{\text s}=&\int_{0}^{\infty}\int_{y}^{\infty}\left[{\mathcal I}_M\left(x\right)-{\mathcal I}_M\left(y\right)\right]f_{\text{B}}\left(x\right)f_{\text{E}}\left(y\right){\rm{d}}x{\rm{d}}y\label{EQUATION8}\\
=&\int_{0}^{\infty}\int_{y}^{\infty}\left[{\mathcal L}_M\left(y\right)-{\mathcal L}_M\left(x\right)\right]f_{\text{B}}\left(x\right)f_{\text{E}}\left(y\right){\rm{d}}x{\rm{d}}y,\label{EQUATION9}
\end{align}
where \eqref{EQUATION9} holds for ${\mathcal I}_M\left(\gamma\right)=\log_2\frac{M}{{\rm e}}-{\mathcal L}_M\left(\gamma\right)$.
{To facilitate the derivation, similar as \cite[eq. (6)]{b7}, we rewrite \eqref{EQUATION9} as ${\bar{{\mathcal I}}}_{\text s}={\mathcal{I}}_3-{\mathcal{I}}_2-{\mathcal{I}}_1$, where ${\mathcal{I}}_3=\int_{0}^{+\infty}{\mathcal L}_M\left(y\right)f_{\text{E}}\left(y\right){\rm{d}}y$, ${\mathcal{I}}_2=\int_{0}^{+\infty}{\mathcal L}_M\left(y\right)f_{\text{E}}\left(y\right)F_{\text{B}}\left(y\right){\rm{d}}y$, and ${\mathcal{I}}_1=\int_{0}^{+\infty}{\mathcal L}_M\left(x\right)f_{\text{B}}\left(x\right)F_{\text{E}}\left(x\right){\rm{d}}x$.} Then it follows that
\begin{align}
&{\mathcal{I}}_1=\sum_{l=1}^{L_{\text B}}\sum_{j=1}^{L_{\text E}}\frac{\alpha_{{\text B},l}\alpha_{{\text E},j}}{\zeta_{{\text E},j}^{\beta_{{\text E},j}}}\int_{0}^{\infty}\frac{{\mathcal L}_M\left(x\right)\Upsilon\left(\beta_{{\text E},j},\zeta_{{\text E},j}x\right)}{x^{1-\beta_{{\text B},l}}{\rm e}^{\zeta_{{\text B},l}x}}{\rm d}x,\\
&{\mathcal{I}}_2=\sum_{l=1}^{L_{\text B}}\sum_{j=1}^{L_{\text E}}\frac{\alpha_{{\text B},l}\alpha_{{\text E},j}}{\zeta_{{\text B},l}^{\beta_{{\text B},l}}}\int_{0}^{\infty}\frac{{\mathcal L}_M\left(y\right)\Upsilon\left(\beta_{{\text B},l},\zeta_{{\text B},l}y\right)}{y^{1-\beta_{{\text E},j}}{\rm e}^{\zeta_{{\text E},j}y}}{\rm d}y,\\
&{\mathcal{I}}_3=\sum\nolimits_{j=1}^{L_{\text E}}\alpha_{{\text E},j}\int_{0}^{\infty}{\mathcal L}_M\left(y\right)y^{\beta_{{\text E},j}-1}{\rm e}^{-\zeta_{{\text E},j}y}{\rm d}y.
\end{align}
We find that ${\mathcal{I}}_1$, ${\mathcal{I}}_2$, and ${\mathcal{I}}_3$ can be efficiently calculated using the Gauss–Laguerre quadrature rule \cite[eq. (25.4.45)]{b14}. Taken together, the final result of ${\bar{{\mathcal I}}}_{\text s}$ is summarized in \eqref{EQUATION13} as follows
\begin{align}
{\bar{{\mathcal I}}}_{{\text s}}&\approx\sum_{j=1}^{L_{\text E}}\frac{\alpha_{{\text E},j}}{\zeta_{{\text E},j}^{\beta_{{\text E},j}}}\sum_{q=1}^{p}\varpi_q{\mathcal L}_{M}\left(\frac{\tau_q}{\zeta_{{\text E},j}}\right)\tau_q^{\beta_{{\text E},j}-1}-
\sum_{l=1}^{L_{\text B}}\sum_{j=1}^{L_{\text E}}\frac{\alpha_{{\text E},j}}{\zeta_{{\text B},l}^{\beta_{{\text B},l}}}\nonumber\\
&\times\frac{\alpha_{{\text E},j}}{\zeta_{{\text E},j}^{\beta_{{\text E},j}}}\sum_{q=1}^{p}\varpi_q\bigg[{\mathcal L}_{M}\left(\frac{\tau_q}{\zeta_{{\text E},j}}\right)\tau_q^{\beta_{{\text E},j}-1}\Upsilon\left(\beta_{{\text B},l},\zeta_{{\text B},l}\frac{\tau_q}{\zeta_{{\text E},j}}\right)\nonumber\\
&+{\mathcal L}_{M}\left(\frac{\tau_q}{\zeta_{{\text B},l}}\right)\tau_q^{\beta_{{\text B},l}-1}\Upsilon\left(\beta_{{\text E},j},\zeta_{{\text E},j}\frac{\tau_q}{\zeta_{{\text B},l}}\right)\bigg]={\hat{{\mathcal I}}}_{{\text s}}^{\left(p\right)},
\label{EQUATION13}
\end{align}
where $\left\{\varpi_q\right\}$ and $\left\{\tau_q\right\}$ denote the weight and abscissas factors of the Gauss–Laguerre integration. Note that a larger value of $p$ yields a higher approximation precision. {By simulation, setting $p=30$ can generally ensure $\left|{\bar{{\mathcal I}}}_{{\text s}}-{\hat{{\mathcal I}}}_{{\text s}}^{\left(p\right)}\right|={\mathcal O}\left(10^{-8}\right)$.}
\vspace{-10pt}
\subsection{Asymptotic Analysis}
\label{SECTION3B}
Denote the average SNRs as $\bar\gamma_i={\mathbb E}\left\{\left|h_i\right|^2\right\}$. The following section will discuss the asymptotic ASR in high SNR regimes. Specifically, we set $\bar\gamma_{\text B}$ to infinity while simultaneously fixing $\bar\gamma_{\text E}$. Before deriving the asymptotic ASR, we rewrite ${\bar{{\mathcal I}}}_{\text s}$ as
\begin{align}
&{\bar{{\mathcal I}}}_{\textrm{s}}=\int_{0}^{\infty}{\mathcal I}_{M}\left(\gamma\right)\left[f_{\text B}\left(\gamma\right)F_{\text E}\left(\gamma\right)+f_{\text E}\left(\gamma\right)F_{\text B}\left(\gamma\right)\right]{\rm d}\gamma\nonumber\\
&-\int_{0}^{\infty}{\mathcal I}_{M}\left(\gamma\right)f_{\text E}\left(\gamma\right){\rm d}\gamma=-\int_{0}^{\infty}{\mathcal I}_{M}\left(\gamma\right)f_{\text E}\left(\gamma\right){\rm d}\gamma\nonumber\\
&+\int_{0}^{\infty}{\mathcal I}_{M}\left(\gamma\right){\rm d}\left(F_{\text B}\left(\gamma\right)F_{\text E}\left(\gamma\right)\right)=\log_2{M}-\int_{0}^{\infty}F_{\text B}\left(\gamma\right)\nonumber\\
&\times F_{\text E}\left(\gamma\right){\rm d}{\mathcal I}_{M}\left(\gamma\right)-\int_{0}^{\infty}{\mathcal I}_{M}\left(\gamma\right)f_{\text E}\left(\gamma\right){\rm d}\gamma.
\end{align}
We define ${\mathcal I}_{\text{{con}}}\triangleq\int_{0}^{\infty}F_{\text B}\left(\gamma\right)F_{\text E}\left(\gamma\right){\rm d}{\mathcal I}_{M}\left(\gamma\right)$ and ${\mathcal I}_{\text{lim}}\triangleq\log_2{M}-\int_{0}^{\infty}{\mathcal I}_{M}\left(\gamma\right)f_{\text E}\left(\gamma\right){\rm d}\gamma$. Note that ${\mathcal I}_{\text{lim}}$ can be calculated by the Gauss–Laguerre quadrature rule, which yields
\begin{align}
{\mathcal I}_{\text{lim}}=\log_2{{\rm e}}+\sum_{j=1}^{L_{\text E}}\frac{\alpha_{{\text E},j}}{\zeta_{{\text E},j}^{\beta_{{\text E},j}}}\sum_{q=1}^{p}\varpi_q{\mathcal L}_{M}\left(\frac{\tau_q}{\zeta_{{\text E},j}}\right)\tau_q^{\beta_{{\text E},j}-1}.
\end{align}

Based on \cite{b8}, $\lim_{\bar\gamma_{\text B}\rightarrow\infty}\zeta_{{\text B},l}=0$ holds, which together with the fact of $\lim_{x\rightarrow0}\Upsilon\left(s,x\right)=\frac{x^s}{s}$ \cite[eq. (8.354.1)]{b15}, yields
\begin{equation}
\lim_{\bar\gamma_{\text B}\rightarrow\infty}F_{\text B}\left(\gamma\right)=\sum_{l=1}^{L_{\text B}}\frac{\alpha_{{\text B},l}}{\zeta_{{\text B},l}^{\beta_{{\text B},l}}}\frac{\left(\zeta_{{\text B},l}\gamma\right)^{\beta_{{\text B},l}}}{\beta_{{\text B},l}}=\sum_{l=1}^{L_{\text B}}\frac{\alpha_{{\text B},l}}{\beta_{{\text B},l}}\gamma^{\beta_{{\text B},l}}.\label{EQUATION17_NEW}
\end{equation}
As explained earlier, the MGD model is proposed to characterize several fading distributions and the values of $L_i$, $\alpha_{i,l}$, $\beta_{i,l}$, $\zeta_{i,l}$ vary with the target distributions. Fortunately, we find that regardless of the target fading distributions, $\sum_{l=1}^{L_{\text B}}\frac{\alpha_{{\text B},l}}{\beta_{{\text B},l}}\gamma^{\beta_{{\text B},l}}$ can be re-expressed as follows
\begin{align}
\sum\nolimits_{l=1}^{L_{\text B}}{\alpha_{{\text B},l}}{\beta_{{\text B},l}^{-1}}\gamma^{\beta_{{\text B},l}}=\sum\nolimits_{l=1}^{L_{\text B}}\Phi_{{\text B},l}\gamma^{\Lambda_{{\text B},l}}{\bar\gamma_{{\text B}}}^{-\Psi_{{\text B},l}},\label{Asym_CDF}
\end{align}
where $\Lambda_{{\text B},l}$, $\Phi_{{\text B},l}$, $\Psi_{{\text B},l}$ are target-distribution-specific constants; $\Lambda_{{\text B},L_{\text B}}\geq\cdots\geq\Lambda_{{\text B},2}\geq\Lambda_{{\text B},1}>0$, $\Phi_{{\text B},l}>0$, and $\Psi_{{\text B},L_{\text B}}\geq\cdots\geq\Psi_{{\text B},2}\geq\Psi_{{\text B},1}>0$. {We comment that \eqref{Asym_CDF} can be directly derived by leveraging the analytical results in \cite{b8} and \cite{b13}. Since the derivation is trivial, we omit the detailed steps here.} Additionally, $\frac{{\rm d}{\mathcal I}_M\left(\gamma\right)}{{\rm d}\gamma}={\mathsf{MMSE}}_M\left(\gamma\right)$ holds \cite{b16}, where ${\mathsf{MMSE}}_M\left(\cdot\right)$ denotes the minimum mean square error (MMSE) function of $M$-QAM. Hence, we obtain
\begin{align}
&\lim\limits_{\bar\gamma_{\text B}\rightarrow\infty}{\mathcal I}_{\text{con}}=\sum\nolimits_{l=1}^{L_{\text B}}\frac{\Phi_{{\text B},l}}{{\bar\gamma_{\text B}}^{\Psi_{{\text B},l}}}\int_{0}^{\infty}\gamma^{\Lambda_{{\text B},l}}F_{\text E}\left(\gamma\right){\mathsf{MMSE}}_M\left(\gamma\right){\rm d}\gamma\nonumber\\
&\quad=\sum\nolimits_{l=1}^{L_{\text B}}\frac{\Phi_{{\text B},l}}{{\bar\gamma_{\text B}}^{\Psi_{{\text B},l}}}{\mathcal M}\left[F_{\text E}\left(\gamma\right){\mathsf{MMSE}}_M\left(\gamma\right);\Lambda_{{\text B},l}+1\right],
\end{align}
where ${\mathcal M}\left[f\left(t\right);z\right]\triangleq\int_{0}^{\infty}t^{z-1}f\left(t\right){\rm d}t$ denotes the Mellin transform of $f\left(t\right)$. For the sake of brevity, define $\Theta_{M,l}\triangleq{\mathcal M}\left[F_{\text E}\left(\gamma\right){\mathsf{MMSE}}_M\left(\gamma\right);\Lambda_{{\text B},l}+1\right]$. Specifically, the following theorem can be found \cite{b18}:
\vspace{-5pt}
\begin{theorem}
\label{theorem1}
  If $f\left(t\right)$ is ${\mathcal O}\left(t^a\right)$ as $t\rightarrow0^{+}$ and ${\mathcal O}\left(t^b\right)$ as $t\rightarrow+\infty$, then ${\mathcal M}\left[f\left(t\right);z\right]$ converges absolutely in the strip $\left\langle-a,-b\right\rangle$ or in other words, $\left|{\mathcal M}\left[f\left(t\right);z\right]\right|<\infty$ when $-a<\Re\left(z\right)<-b$. Here, $\Re\left(z\right)$ denotes the real part of $z$
\end{theorem}
\vspace{-5pt}
Based on \cite{b19}, we have $\lim_{\gamma\rightarrow0^{+}}F_{\text E}\left(\gamma\right){\mathsf{MMSE}}_M\left(\gamma\right)=0$ and $\lim_{\gamma\rightarrow\infty}F_{\text E}\left(\gamma\right){\mathsf{MMSE}}_M\left(\gamma\right)=o\left({\rm e}^{-d_M\gamma}\right)$ ($d_M>0$). Therefore, $F_{\text E}\left(\gamma\right){\mathsf{MMSE}}_M\left(\gamma\right)$ is ${\mathcal O}\left(\gamma^k\right)$ ($k>0$) as $\gamma\rightarrow0^{+}$ and ${\mathcal O}\left(x^l\right)$ ($l=-\infty$) as $x\rightarrow+\infty$. As stated before, $\Lambda_{{\text B},l}>0$; thus, $\Re\left(\Lambda_{{\text B},l}+1\right)\in\left(-k,-l\right)$, which together with Theorem \ref{theorem1}, yields $\left|\Theta_{M,l}\right|<\infty$. Moreover, by \cite{b16}, ${\mathsf{MMSE}}_M\left(\gamma\right)>0$, which yields $\Theta_{M,l}>0$. In summary, $\forall l\in[1,L_{\text B}]$, $\Theta_{M,l}\in\left(0,+\infty\right)$ and $\Theta_{M,l}$ can be calculated by methods of numerical integration. Hence, we obtain
\begin{equation}
\lim\nolimits_{\bar\gamma_{\text B}\rightarrow\infty}{\mathcal I}_{\text{con}}={G_{a,{\text B},M}}{{\bar\gamma_{\text B}}^{-G_{d,{\text B}}}}+{o}\left({{\bar\gamma_{\text B}}^{-G_{d,{\text B}}}}\right),
\end{equation}
where $G_{a,{\text B},M}=\sum_{\Psi_{{\text B},l}=\Psi_{{\text B},1}}\Theta_{M,l}\Phi_{{\text B},l}>0$ and $G_{d,{\text B}}=\Psi_{{\text B},1}>0$; $o\left(\cdot\right)$ denotes the higher order term. Consequently, the asymptotic ASR can be written as
\begin{equation}
{\bar{{\mathcal I}}}_{s}^{\infty}={\mathcal I}_{\text{lim}}-{G_{a,{\text B},M}}{{\bar\gamma_{\text B}}^{-G_{d,{\text B}}}}+{o}\left({{\bar\gamma_{\text B}}^{-G_{d,{\text B}}}}\right).\label{EQUATION17}
\end{equation}
Based on \eqref{EQUATION17}, the ASR converges to ${\mathcal I}_{\text{lim}}$ as $\bar\gamma_{\text B}\rightarrow\infty$ and the rate of convergence (ROC) is determined by ${G_{a,{\text B},M}}$ and $G_{d,{\text B}}$. Besides, it can be concluded from \eqref{EQUATION17} that the secrecy diversity order (SDO) is $G_{d,{\text B}}=\Psi_{{\text B},1}$. {Actually, the fact of $\lim_{\bar\gamma_{\text B}\rightarrow\infty}{\bar{\mathcal I}}_{\text s}={\mathcal I}_{\text{lim}}$ was observed in \cite{b11} and the references therein, but the ROC of ${\bar{\mathcal I}}_{\text s}$ has not been investigated in prior works. Last but not least, aided with \cite{b8,b13}, we obtain $\lim_{\bar\gamma_i\rightarrow\infty}f_i\left(\gamma_i\right)=\sum_{l=1}^{L_{i}}\Phi_{{i},l}{\Lambda_{{i},l}}\gamma_i^{\Lambda_{{i},l}-1}{\bar\gamma_{{i}}}^{-\Psi_{{i},l}}$. Based on this and \eqref{Asym_CDF}, the asymptotic ASR when both $\bar\gamma_{\text B}$ and $\bar\gamma_{\text E}$ approach infinity can be also characterized. Since the derivations are similar as those above, the details are left to our future work.}
\vspace{-5pt}
\section{Secrecy Outage Probability}
\subsection{Explicit Analysis}
The SOP is defined as the probability when the instantaneous secrecy rate is lower than a preset value $R_{\text s}>0$, and it is written as $P\left(R_{\text s}\right)=\Pr\left({\mathcal I}_{M}^{\text{s}}<R_{\text s}\right)$. By \eqref{EQUATION4}, when $\gamma_{\text B}<\gamma_{\text E}$, ${\mathcal I}_{M}^{\text{s}}=0<R_{\text s}$; thus, $P\left(R_{\text s}\right)$ can be further written as
\begin{equation}
P\left(R_{\text s}\right)=\Pr\left({\mathcal I}_{M}^{\text{s}}<R_{\text s},\gamma_{\text B}>\gamma_{\text E}\right)+\Pr\left(\gamma_{\text B}<\gamma_{\text E}\right).\label{EQUATION18}
\end{equation}
Define ${\mathcal H}_M\triangleq {\mathcal I}_M^{-1}\left(\log_2 M-R_{\text s}\right)$ and ${\mathcal F}_M\left(\gamma\right)\triangleq{{\mathcal I}_M^{-1}\left(R_{\text s}+{\mathcal I}_M\left(\gamma\right)\right)}$, where ${\mathcal I}_{M}^{-1}\left(\cdot\right)$ denotes the inverse function of ${\mathcal I}_{M}\left(\cdot\right)$. Though ${\mathcal I}_{M}^{-1}\left(\cdot\right)$ lacks an explicit expression, its value can be found via a simple bisection search. Then, the following theorem can be found:
\vspace{-5pt}
\begin{theorem}
\label{theorem2}
  The SOP with discrete inputs is given by
  \vspace{-5pt}
  \begin{equation}
  P\left(R_{\rm s}\right)=1-F_{\rm{E}}\left({\mathcal H}_M\right)+\int_{0}^{{ {\mathcal H}_M}}F_{\rm{B}}\left({{\mathcal F}_M\left(y\right)}\right)f_{\rm{E}}\left(y\right){\rm{d}}y.\label{EQUATION26}
  \end{equation}
\end{theorem}
\vspace{-5pt}
\begin{IEEEproof}
Please see Appendix \ref{Append1}.
\end{IEEEproof}
Substituting \eqref{EQUATION6} and \eqref{EQUATION7} into \eqref{EQUATION26} gives
\begin{align}
P\left(R_{\text s}\right)&=1-\sum_{l=1}^{L_{\text E}}\frac{\alpha_{{\text E},l}}{\zeta_{{\text E},l}^{\beta_{{\text E},l}}}\Upsilon\left(\beta_{{\text E},l},\zeta_{{\text E},l}{{\mathcal H}_M}\right)+
\sum_{q=1}^{L_{\text B}}\sum_{j=1}^{L_{\text E}}\frac{\alpha_{{\text B},q}\alpha_{{\text E},j}}{\zeta_{{\text B},q}^{\beta_{{\text B},q}}}\nonumber\\
&\times\int_{0}^{{\mathcal H}_M}{\Upsilon\left(\beta_{{\text B},q},\zeta_{{\text B},q}{{\mathcal F}_M\left(y\right)}\right)}y^{\beta_{{\text E},j}-1}{\rm e}^{-\zeta_{{\text E},j}y}{\rm{d}}y.\label{EQUATION27}
\end{align}
The integral in \eqref{EQUATION27} can be effectively evaluated by the Gauss-Legendre quadrature rule \cite[eq. (25.4.30)]{b14}, which yields
\begin{align}
&P\left(R_{\text s}\right)\approx1-\sum_{l=1}^{L_{\text E}}\frac{\alpha_{{\text E},l}}{\zeta_{{\text E},l}^{\beta_{{\text E},l}}}\Upsilon\left(\beta_{{\text E},l},\zeta_{{\text E},l}{{\mathcal H}_M}\right)+
\sum_{q=1}^{L_{\text B}}\sum_{j=1}^{L_{\text E}}\frac{\alpha_{{\text B},q}\alpha_{{\text E},j}}{2\zeta_{{\text B},q}^{\beta_{{\text B},q}}}\nonumber\\
&\times\sum_{i=1}^{v}\frac{{{\mathcal H}_M}\vartheta_i\Upsilon\left(\beta_{{\text B},q},\zeta_{{\text B},q}{{\mathcal F}_M\left(\frac{{\mathcal H}_M}{2}\xi_i+\frac{{\mathcal H}_M}{2}\right)}\right)}{\left(\frac{{\mathcal H}_M}{2}\xi_i+\frac{{\mathcal H}_M}{2}\right)^{1-\beta_{{\text E},j}}\exp\left(\frac{{\mathcal H}_M\zeta_{{\text E},j}}{2}\left(\xi_i+1\right)\right)}.\label{EQUATION28}
\end{align}
Here, $\left\{\vartheta_i\right\}$ and $\left\{\xi_i\right\}$ denote the weight and abscissas factors of the Gauss-Legendre integration. Note that a larger value of $v$ yields a higher approximation precision. {We find that approximately $v=30$ is required to achieve $10^{-7}$ accuracy.}
\vspace{-5pt}
\subsection{Asymptotic Analysis}
\label{SECTION4B}
Based on \eqref{EQUATION17_NEW} and \eqref{Asym_CDF}, when $\gamma$ is fixed, $\lim_{{\bar\gamma_{\text B}}\rightarrow\infty}F_{\text B}\left(\gamma\right)=\sum_{l=1}^{L_{\text B}}\Phi_{{\text B},l}\gamma^{\Lambda_{{\text B},l}}{\bar\gamma_{{\text B}}}^{-\Psi_{{\text B},l}}$. Thus, as ${\bar\gamma_{\text B}}\rightarrow\infty$, \eqref{EQUATION26} can be written as $P_{\text{out}}^{\infty}=\lim_{{\bar\gamma_{\text B}}\rightarrow\infty}P\left(R_{\rm s}\right)=1-F_{\text{E}}\left({\mathcal H}_M\right)+\sum_{l=1}^{L_{\text B}}\Delta_{M,l}\frac{\Phi_{{\text B},l}}{{\bar\gamma_{\text B}^{\Psi_{{\text B},l}}}}$, where $\Delta_{M,l}=\int_{0}^{{\mathcal H}_M}{\mathcal F}_M^{\Lambda_{{\text B},l}}\left(\gamma\right)f_{\text{E}}\left(\gamma\right){\rm{d}}\gamma$. By using variable substitution $\gamma\rightarrow {\mathcal Z}_{M}\left(x\right)={\mathcal I}_{M}^{-1}\left({\mathcal I}_M\left(x\right)-R_{\text s}\right)$, we obtain
\begin{equation}
\Delta_{M,l}=\int_{{\mathcal I}_{M}^{-1}\left(R_{\text s}\right)}^{+\infty}x^{\Lambda_{{\text B},l}}\frac{f_{\text E}\left({\mathcal Z}_{M}\left(x\right)\right){\mathsf{MMSE}}_M\left(x\right)}{{\mathsf{MMSE}}_M\left({\mathcal Z}_{M}\left(x\right)\right)}{\rm d}x.
\end{equation}
For simplicity, we define ${\mathcal W}_M\left(x\right)\triangleq\frac{f_{\text E}\left({\mathcal Z}_{M}\left(x\right)\right){\mathsf{MMSE}}_M\left(x\right)}{{\mathsf{MMSE}}_M\left({\mathcal Z}_{M}\left(x\right)\right)}$ and
\begin{equation}
{\mathcal K}_M\left(x\right)\triangleq\left\{
\begin{array}{lr}
{\mathcal W}_M\left({\mathcal I}_{M}^{-1}\left(R_{\text s}\right)\right)=0,&{0<x\leq{\mathcal I}_{M}^{-1}\left(R_{\text s}\right)}\\
{\mathcal W}_M\left(x\right),&{x>{\mathcal I}_{M}^{-1}\left(R_{\text s}\right)}
\end{array}\right..\nonumber
\end{equation}
Hence, we have $\Delta_{M,l}={\mathcal M}\left[{\mathcal K}_M\left(x\right);\Lambda_{{\text B},l}+1\right]$. Particularly, the following theorem captures the main result of the asymptotic SOP.
\vspace{-5pt}
\begin{theorem}
\label{theorem3}
  The asymptotic SOP can be written as
  \vspace{-5pt}
  \begin{equation}
P_{\textrm{out}}^{\infty}=1-F_{\text{E}}\left({\mathcal H}_M\right)+{G'_{a,{\text B},M}}{{\bar\gamma_{\text B}}^{-G_{d,{\text B}}}}+{o}\left({{\bar\gamma_{\text B}}^{-G_{d,{\text B}}}}\right),\label{EQUATION35}
\end{equation}
where $G'_{a,{\text B},M}=\sum_{\Psi_{{\text B},l}=\Psi_{{\text B},1}}\Delta_{M,l}\Phi_{{\text B},l}>0$ and $G_{d,{\text B}}=\Psi_{{\text B},1}>0$.
\end{theorem}
\vspace{-5pt}
\begin{IEEEproof}
Please see Appendix \ref{Append2}.
\end{IEEEproof}
The proof in Appendix \ref{Append2} suggests that $\Delta_{M,l}\in\left(0,\infty\right)$ and $\Delta_{M,l}$ can be calculated by the Gauss-Legendre quadrature rule. Based on \eqref{EQUATION35}, we find that the SOP converges to $1-F_{\text{E}}\left({\mathcal H}_M\right)$ as $\bar\gamma_{\text B}\rightarrow\infty$ and the rate of convergence is determined by $G_{d,{\text B}}$ and $G'_{a,{\text B},M}$. Besides, it can be concluded from \eqref{EQUATION35} that the secrecy diversity order is $G_{d,{\text B}}$. We note that the ASR and SOP yield the same SDO but their ROCs are different. {Following the similar steps as described above, the asymptotic SOP when both $\bar\gamma_{\text B}$ and $\bar\gamma_{\text E}$ approach infinity can be also discussed, which will be left to our future work.}
\vspace{-5pt}
\section{Simulation}
\label{sec4}
\begin{table*}[!t]
\centering
\caption{Simulations Parameters \cite{b8}}
\label{TABLE1}
\vspace{-10pt}
\begin{tabular}{c|c|c|c}
\hline
\textbf{Distribution}                   & \textbf{Parameter} & \textbf{Asymptotic}  & $G_{d,\text B}$             \\ \hline
\multicolumn{1}{c|}{Nakagami-$m$ ($m\geq0.5$)} & \multicolumn{1}{c|}{$L_i=1$, $\alpha_{i,1}=\frac{m_i^{m_i}}{\bar\gamma_i^{m_i}\Gamma\left(m_i\right)}$, $\beta_{i,1}=m_i$, $\zeta_{i,1}=\frac{m_i}{\bar\gamma_i}$}  & \multicolumn{1}{c|}{$\Phi_{{\text B},1}=\frac{m_{\text B}^{m_{\text B}-1}}{\Gamma\left(m_{\text B}\right)}$, ${\Lambda_{{\text B},1}}=m_{\text B}$, ${\Psi_{{\text B},1}}=m_{\text B}$}  & $m_{\text B}$                   \\ \hline
\multicolumn{1}{c|}{\multirow{2}{*}{Hoyt (Nakagami-$q$) ($0<q<1$)}} & \multicolumn{1}{c|}{$L_i=20$, $\alpha_{i,l}=\psi\left(\theta_{i,l},\beta_{i,l},\zeta_{i,l}\right)$, $\beta_{i,l}=2l-1$} & \multicolumn{1}{c|}{$\Phi_{{\text B},1}=\frac{q_{\text B}+q_{\text B}^{-1}}{2}$, ${\Lambda_{{\text B},1}}=1$, ${\Psi_{{\text B},1}}=1$,} & \multirow{2}{*}{1} \\
\multicolumn{1}{c|}{}                   & \multicolumn{1}{c|}{$\zeta_{i,l}=\frac{\left(1+q_i^2\right)^2}{4q_i^2\bar\gamma_i}$, $\theta_{i,l}=\frac{\left(1+q_i^2\right)}{2q_i\bar\gamma_i\Gamma\left(l\right)\left(l-1\right)!}\left(\frac{1-q_i^4}{8q_i^2\bar\gamma_i}\right)^{2l-2}$} & \multicolumn{1}{c|}{$\Psi_{{\text B},L_{\text B}}>\cdots>\Psi_{{\text B},2}>\Psi_{{\text B},1}>0$} &                     \\ \hline
\multicolumn{1}{c|}{${\mathcal K}_G$ ($m>0$, $k>0$), $m$ and $k$}                  & \multicolumn{1}{c|}{$L_i=15$, $\alpha_{i,l}=\psi\left(\theta_{i,l},\beta_{i,l},\zeta_{i,l}\right)$, $\beta_{i,l}=m_i$,}  & \multicolumn{1}{c|}{$\Psi_{{\text B},l}=\Psi_{{\text B},1}=m_{\text B}$, $\Lambda_{{\text B},l}=\Lambda_{{\text B},1}=m_{\text B}$,}  & \multirow{2}{*}{$m_{\text B}$}  \\
\multicolumn{1}{c|}{are distribution shaping parameters}                  & \multicolumn{1}{c|}{$\lambda_i=\frac{k_im_i}{\bar\gamma_i}$, $\zeta_{i,l}=\frac{\lambda_i}{t_l}$, $\theta_{i,l}=\frac{\lambda_i^{m_i}\varpi_l\tau_l^{k-m-1}}{\Gamma\left(m_i\right)\Gamma\left(k_i\right)}$}  & \multicolumn{1}{c|}{$\sum\limits_{l=1}^{L_{\text B}}\Phi_{{\text B},l}=\frac{k_{\text B}^{m_{\text B}}m_{\text B}^{m_{\text B}-1}}{\Gamma\left(m_{\text B}\right)\Gamma\left(k_{\text B}\right)}\sum\limits_{l=1}^{L_{\text B}}\varpi_l\tau^{k_{\text B}-m_{\text B}}$}  &                     \\ \hline
\multicolumn{1}{c|}{\multirow{2}{*}{$\kappa$-$\mu$ ($\kappa>0$, $\mu>0$)}} & \multicolumn{1}{c|}{$L_i=20$, $\alpha_{i,l}=\psi\left(\theta_{i,l},\beta_{i,l},\zeta_{i,l}\right)$, $\beta_{i,l}=\mu_i-1+l$,}  & \multicolumn{1}{c|}{$\Phi_{{\text B},1}=\frac{\mu_{\text B}^{\mu_{\text B}-1}\left(1+\kappa_{\text B}\right)^{\mu_{\text B}}}{\Gamma\left(\mu_{\text B}\right)\exp\left(\kappa_{\text B}\mu_{\text B}\right)}$, ${\Lambda_{{\text B},1}}=\mu_{\text B}$,}  & \multirow{2}{*}{$\mu_{\text B}$}  \\
\multicolumn{1}{c|}{}                   & \multicolumn{1}{c|}{$\zeta_{i,l}=\frac{\mu_i\left(1+\kappa_i\right)}{\bar\gamma_i}$, $\theta_{i,l}=\frac{\mu_i^{2l+\mu_i-2}\left(1+\kappa_i\right)^{{\mu_i+l-1}}\kappa_i^{l-1}}{{\rm{e}}^{\mu_i\kappa_i}\bar\gamma_i^{\mu_i+l-1}\Gamma\left(\mu_i-1+l\right)\left(l-1\right)!}$}  & \multicolumn{1}{c|}{$\Psi_{{\text B},L_{\text B}}>\cdots>\Psi_{{\text B},2}>\Psi_{{\text B},1}=\mu_{\text B}$}  &                     \\ \hline
\end{tabular}
\end{table*}

\begin{figure}[!t]
    \centering
    \subfigbottomskip=0pt
	\subfigcapskip=-5pt
\setlength{\abovecaptionskip}{10pt}
    \subfigure[ASR versus $\bar\gamma_{\text B}$, $\bar\gamma_{\text E}=0$ dB.]
    {
        \includegraphics[height=0.185\textwidth]{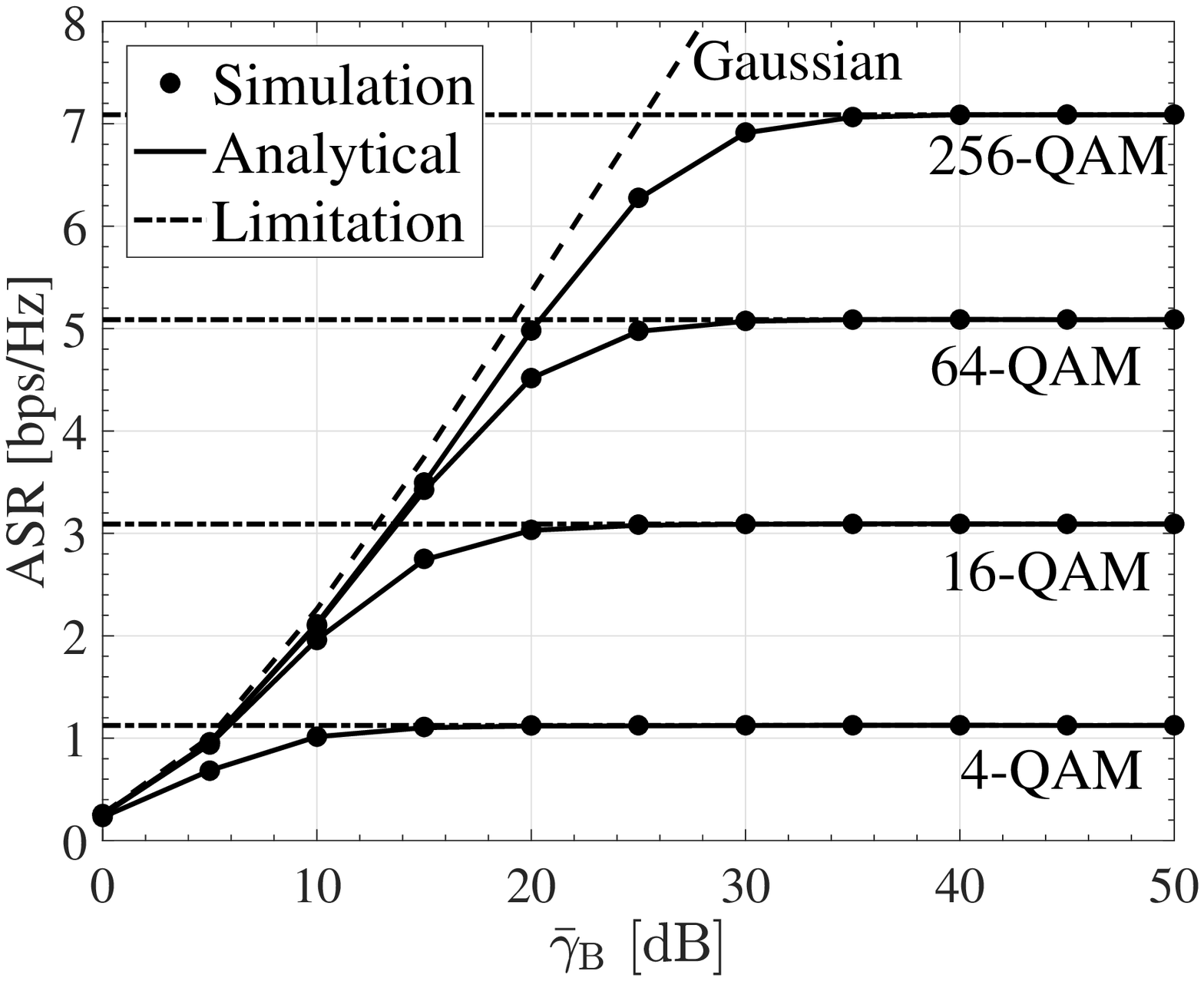}
	   \label{fig1a}	
    }
    \hspace{-14pt}
   \subfigure[${{{\mathcal I}}}_{\text{con}}$ versus $\bar\gamma_{\text B}$, $\bar\gamma_{\text E}=0$ dB.]
    {
        \includegraphics[height=0.185\textwidth]{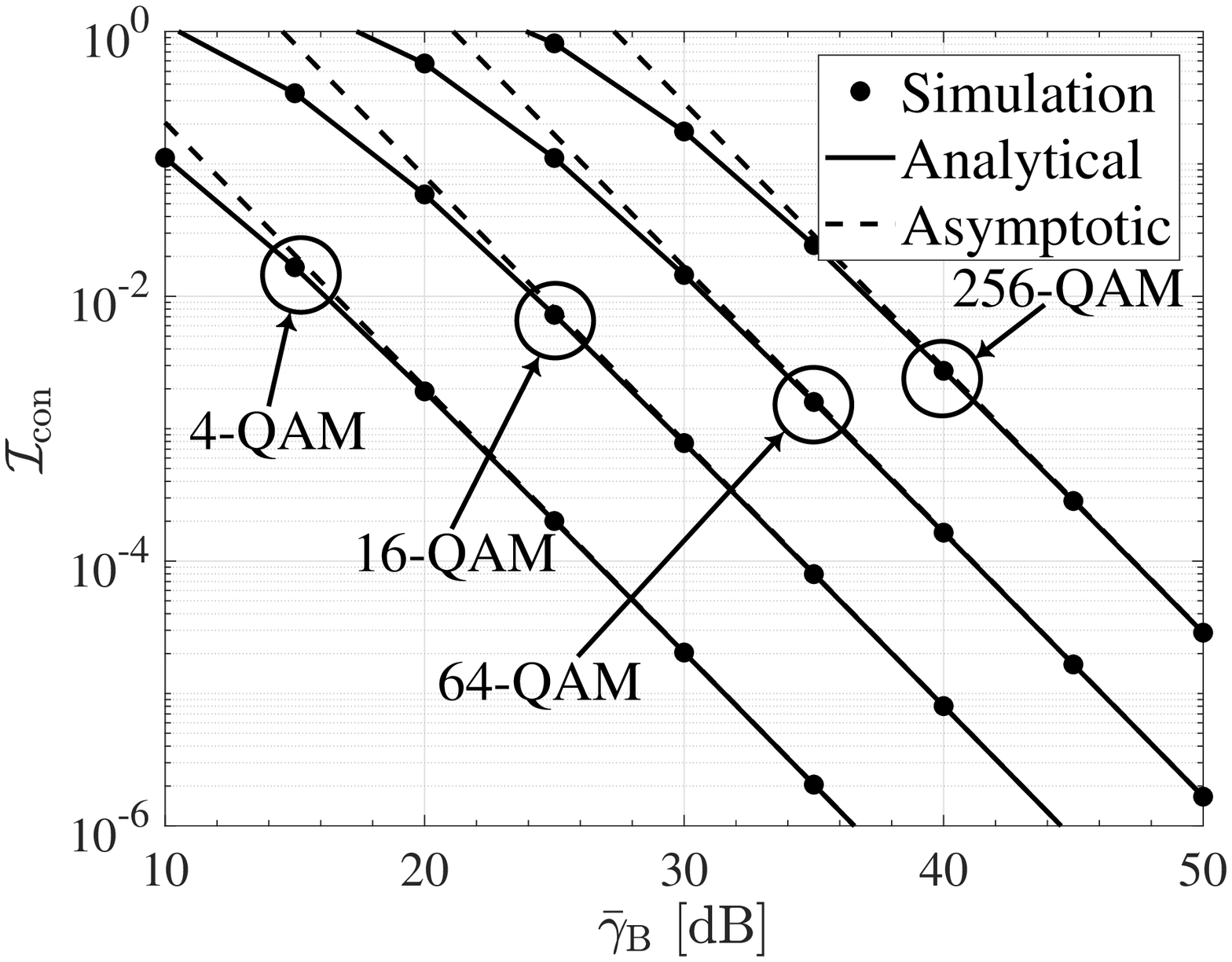}
	   \label{fig1b}	
    }
\caption{ASR versus $\bar\gamma_{\text B}$ over Nakagami-$m$ channels for $m_{\text B}=m_{\text E}=2$.}
    \label{figure1}
\end{figure}

\begin{figure}[!t]
    \centering
    \subfigbottomskip=0pt
	\subfigcapskip=-5pt
\setlength{\abovecaptionskip}{10pt}
    \subfigure[ASR versus $\bar\gamma_{\text B}$.]
    {
        \includegraphics[height=0.179\textwidth]{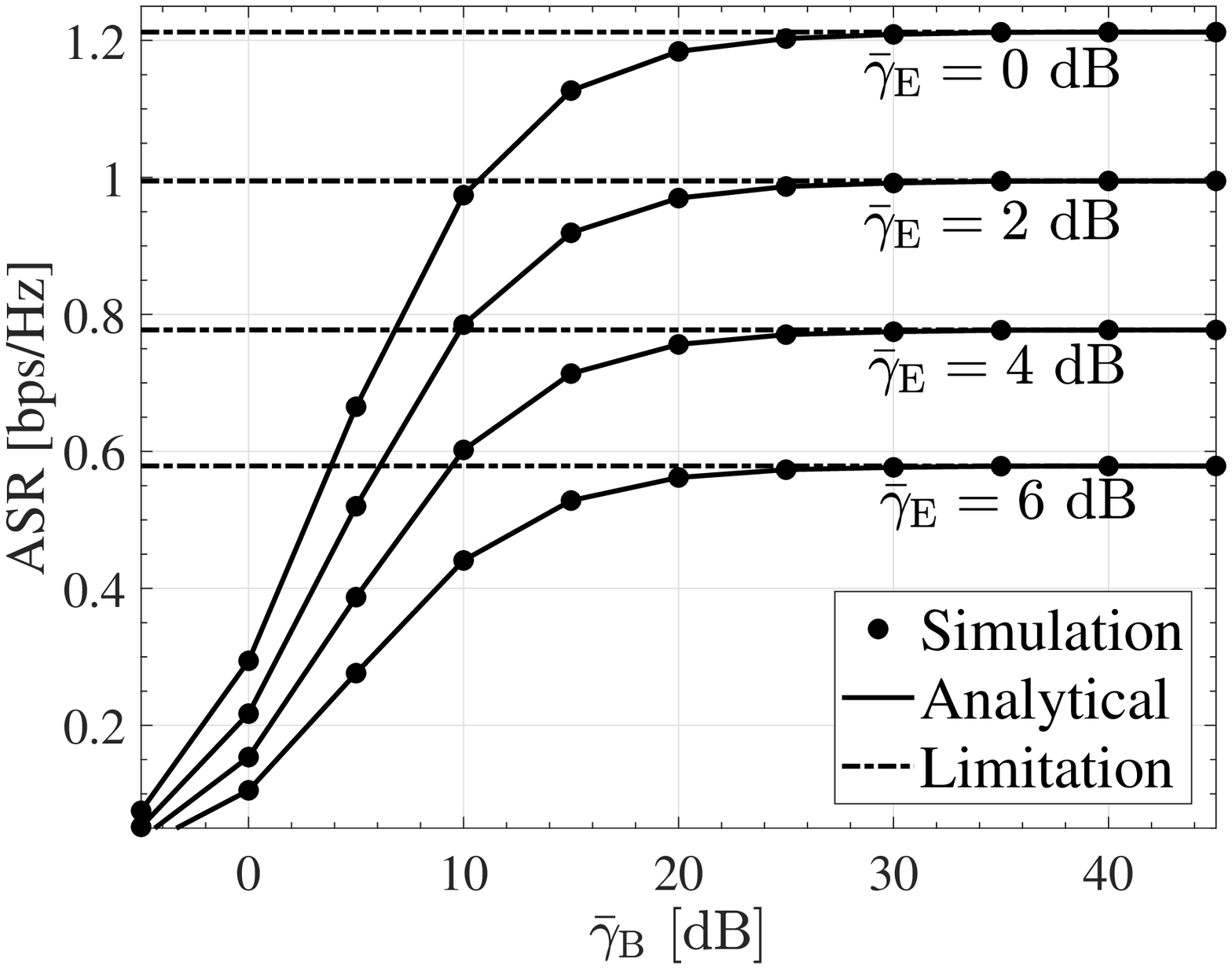}
	   \label{fig2a}	
    }\hspace{-8pt}
   \subfigure[${{{\mathcal I}}}_{\text{con}}$ versus $\bar\gamma_{\text B}$.]
    {
        \includegraphics[height=0.183\textwidth]{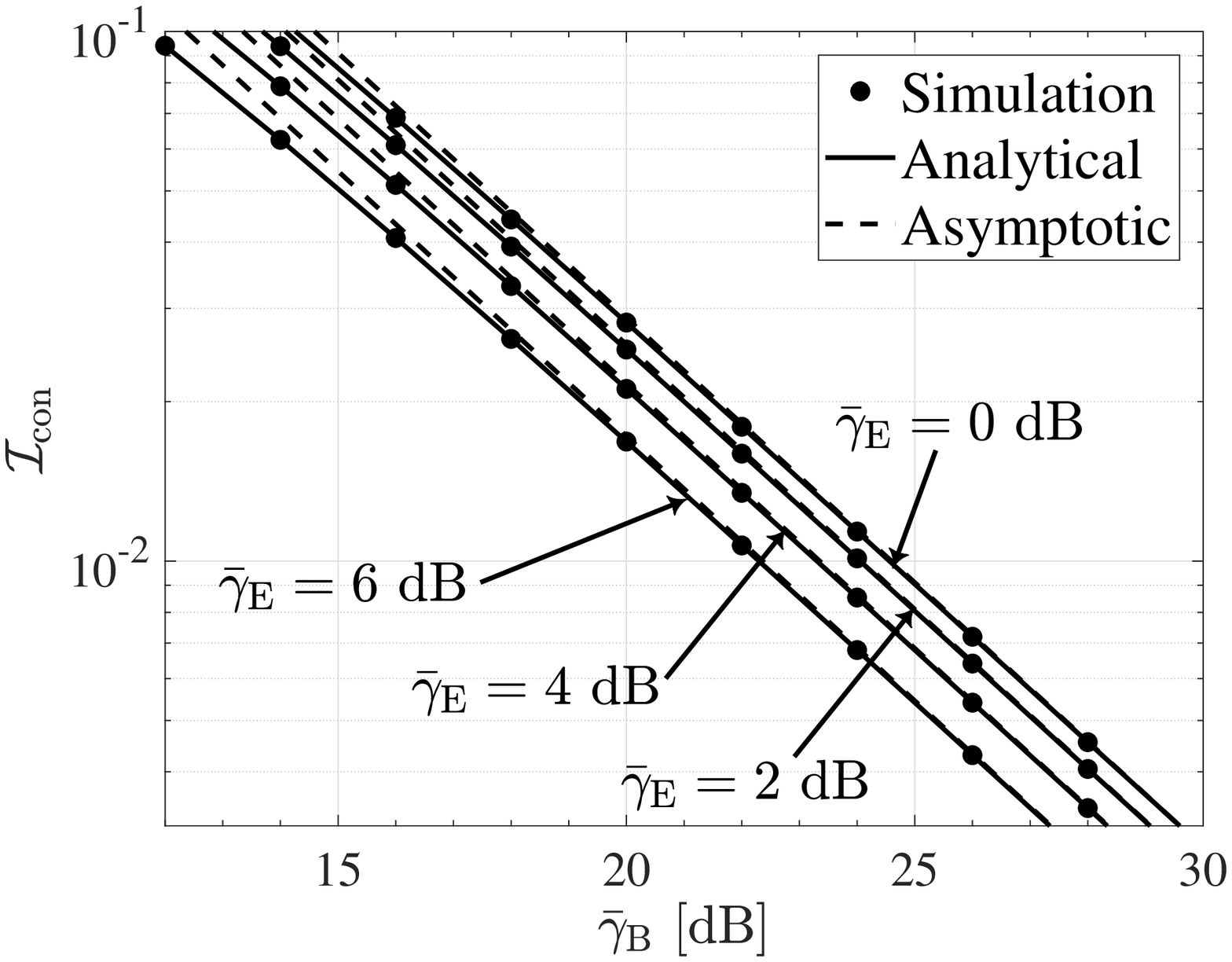}
	   \label{fig2b}	
    }
\caption{ASR of 4-QAM versus $\bar\gamma_{\text B}$ over Hoyt channels for ${q_{\text B}}={q_{\text E}}=\sqrt{0.5}$.}
    \label{figure2}
\end{figure}

As explained earlier, the MGD model serves as a general tool to characterize various types of fading distributions. In this part, we provide four examples listed in Table \ref{TABLE1} to verify our former derivations. In Table \ref{TABLE1}, $\left\{\varpi_q\right\}$ and $\left\{\tau_q\right\}$ denote the weight and abscissas factors of the Gauss–Laguerre integration and $\psi\left(\theta_{i,l},\beta_{i,l},\zeta_{i,l}\right)=\frac{\theta_{i,l}}{\sum_{j=1}^{L_i}\theta_{i,j}\Gamma\left(\beta_{i,j}\right)\zeta_{i,j}^{-\beta_{i,j}}}$.
The simulations are based on the commonly used square $M$-QAM signals.

To validate the precision of \eqref{EQUATION13}, we plot the ASR over Nakagami-$m$ fading channels for various modulation schemes in {\figurename} \ref{fig1a} and compare the analytical results with the simulated results. {The analytical ASR is calculated by \eqref{EQUATION1} and \eqref{EQUATION13}. As stated before, setting $n=20$ can generally ensure ${\epsilon_M^{\left(n\right)}}\left(\gamma\right)={\mathcal O}\left(10^{-5}\right)$ and thus we set $n=20$ to approximate the ASR.} Besides, the value of $p$ in \eqref{EQUATION13} is set as 30. As shown, the analytical results fit well with the simulations. Moreover, we plot the secrecy rate achieved by Gaussian signaling in {\figurename} \ref{figure1} for comparison. As shown, the ASR of Gaussian inputs tends to infinity as $\bar\gamma_{\text B}$ increases, whereas the ASR of discrete inputs converges to its limitation, namely ${\mathcal I}_{\text{lim}}$, in the large limit of $\bar\gamma_{\text B}$. As discussed in Section \ref{SECTION3B}, $\bar{\mathcal I}_{\text s}={\mathcal I}_{\text{con}}+{\mathcal I}_{\text{lim}}$, where $\lim_{\bar\gamma_{\text B}\rightarrow\infty}{\mathcal I}_{\text{con}}=0$ and ${\mathcal I}_{\text{lim}}$ is a constant. This means that the ROC of $\bar{\mathcal I}_{\text s}$ equals that of ${\mathcal I}_{\text{con}}$. To show the ROC of $\bar{\mathcal I}_{\text s}$, we plot ${\mathcal I}_{\text{con}}$ versus $\bar\gamma_{\text B}$ in {\figurename} \ref{fig1b}. As shown, in the high SNR regime, the derived asymptotic results track the numerical results accurately. Besides, it can be observed that a lower modulation order yields a faster ROC. {\figurename} \ref{fig2a} plots the ASR of 4-QAM versus $\bar\gamma_{\text B}$ over Hoyt channels for selected values of $\bar\gamma_{\text E}$. As shown, the ASR decreases with $\bar\gamma_{\text E}$, suggesting the passive influence of the eavesdropper. Then we use {\figurename} \ref{fig2b} to illustrate the ROC of the ASR. It can be seen from this figure that a larger value of $\bar\gamma_{\text E}$ corresponds to a faster ROC.

\begin{figure}[!t]
    \centering
    \subfigbottomskip=0pt
	\subfigcapskip=-5pt
\setlength{\abovecaptionskip}{10pt}
    \subfigure[SOP versus $\bar\gamma_{\text B}$.]
    {
        \includegraphics[height=0.185\textwidth]{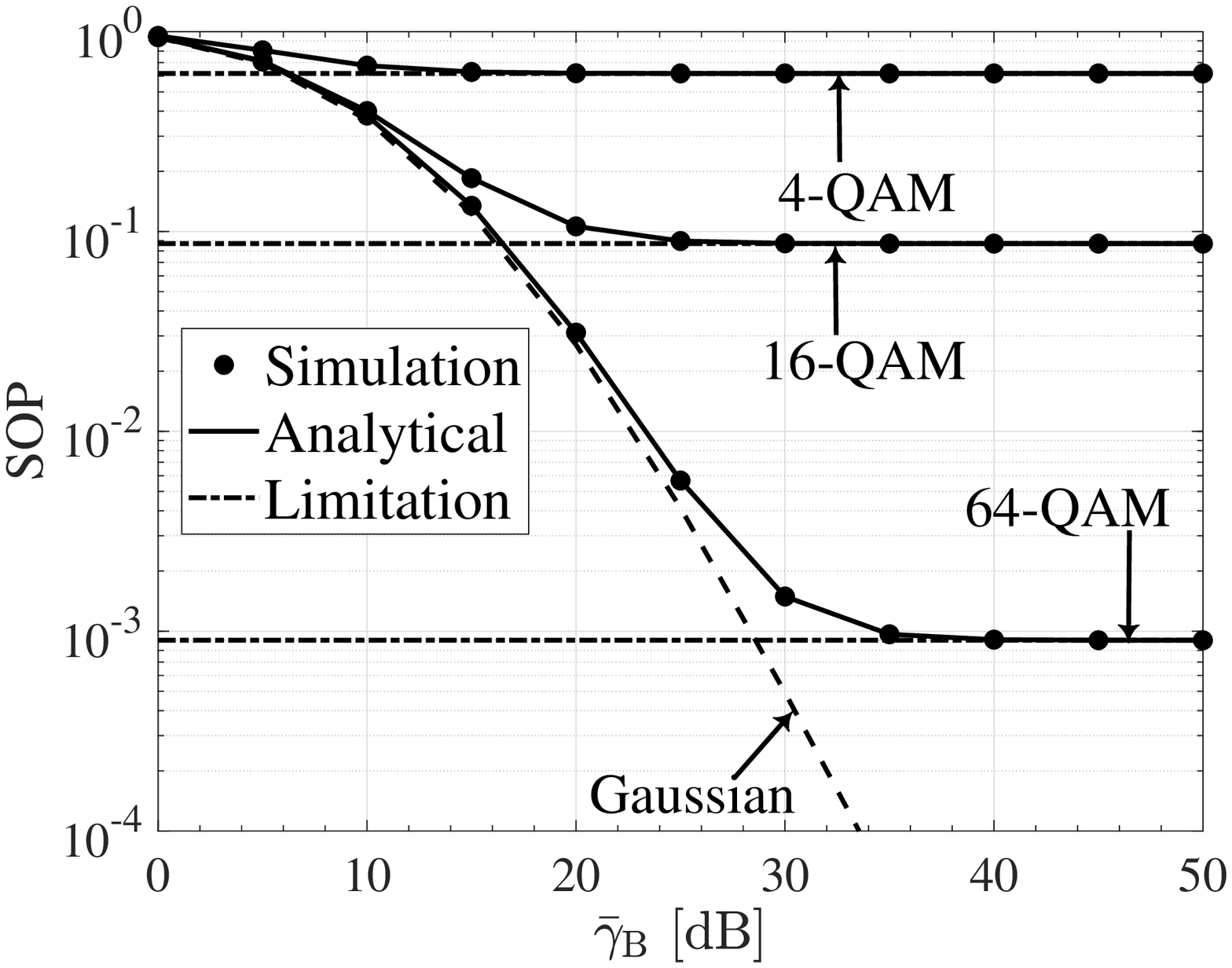}
	   \label{fig3a}	
    }
    \hspace{-14pt}
   \subfigure[$P_{\text{con}}$ versus $\bar\gamma_{\text B}$.]
    {
        \includegraphics[height=0.185\textwidth]{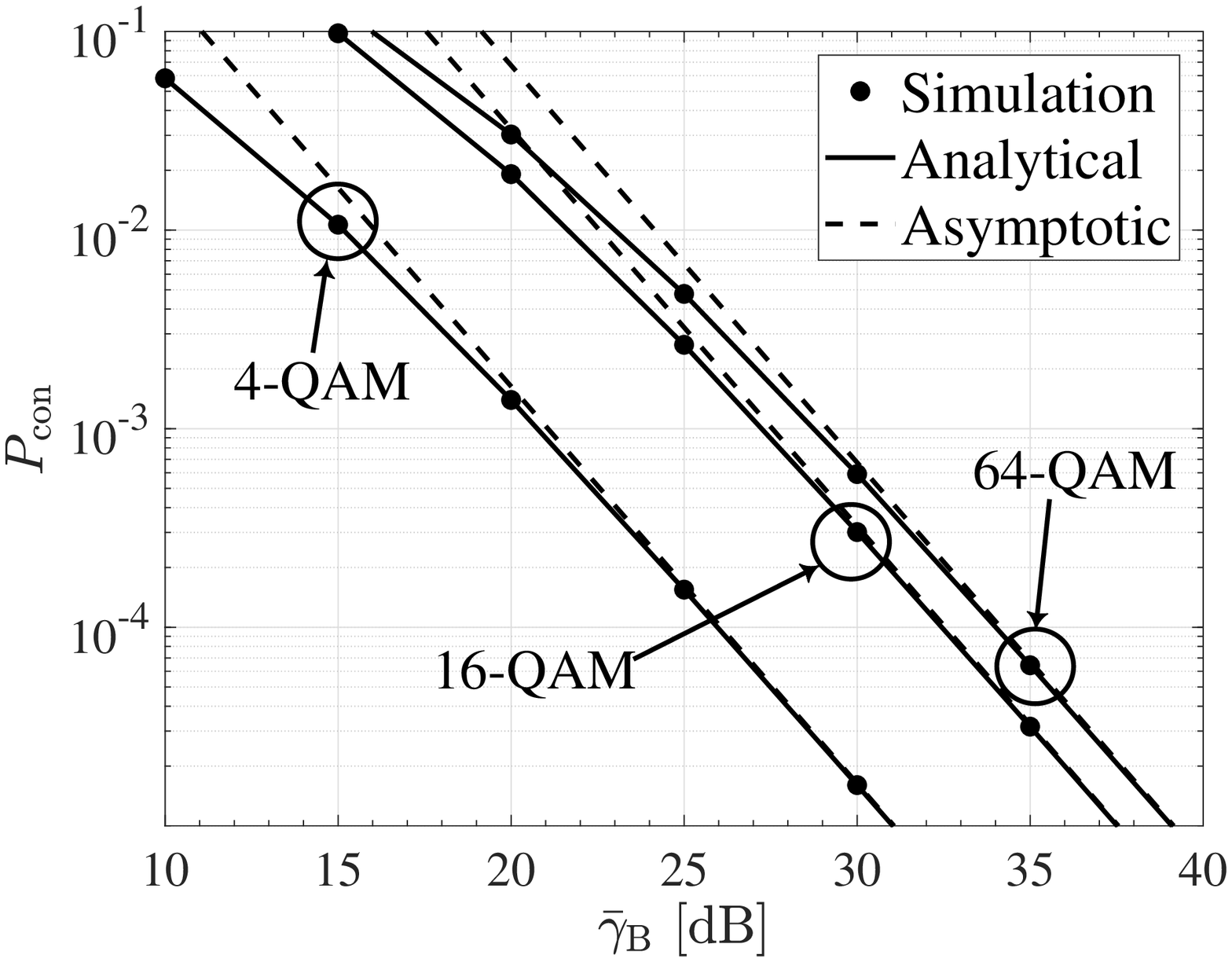}
	   \label{fig3b}	
    }
\caption{SOP of $M$-QAM versus $\bar\gamma_{\text B}$ over ${\mathcal K}_G$ channels for $\bar\gamma_{\text E}=5$ dB, $k_{\text B}=5$, $m_{\text B}=2$, $k_{\text E}=2$, $m_{\text E}=1$, and $R_{\text s}=1$ bps/Hz.}
    \label{figure3}
\end{figure}

\begin{figure}[!t]
    \centering
    \subfigbottomskip=0pt
	\subfigcapskip=-5pt
\setlength{\abovecaptionskip}{10pt}
    \subfigure[SOP versus $\bar\gamma_{\text B}$.]
    {
        \includegraphics[height=0.185\textwidth]{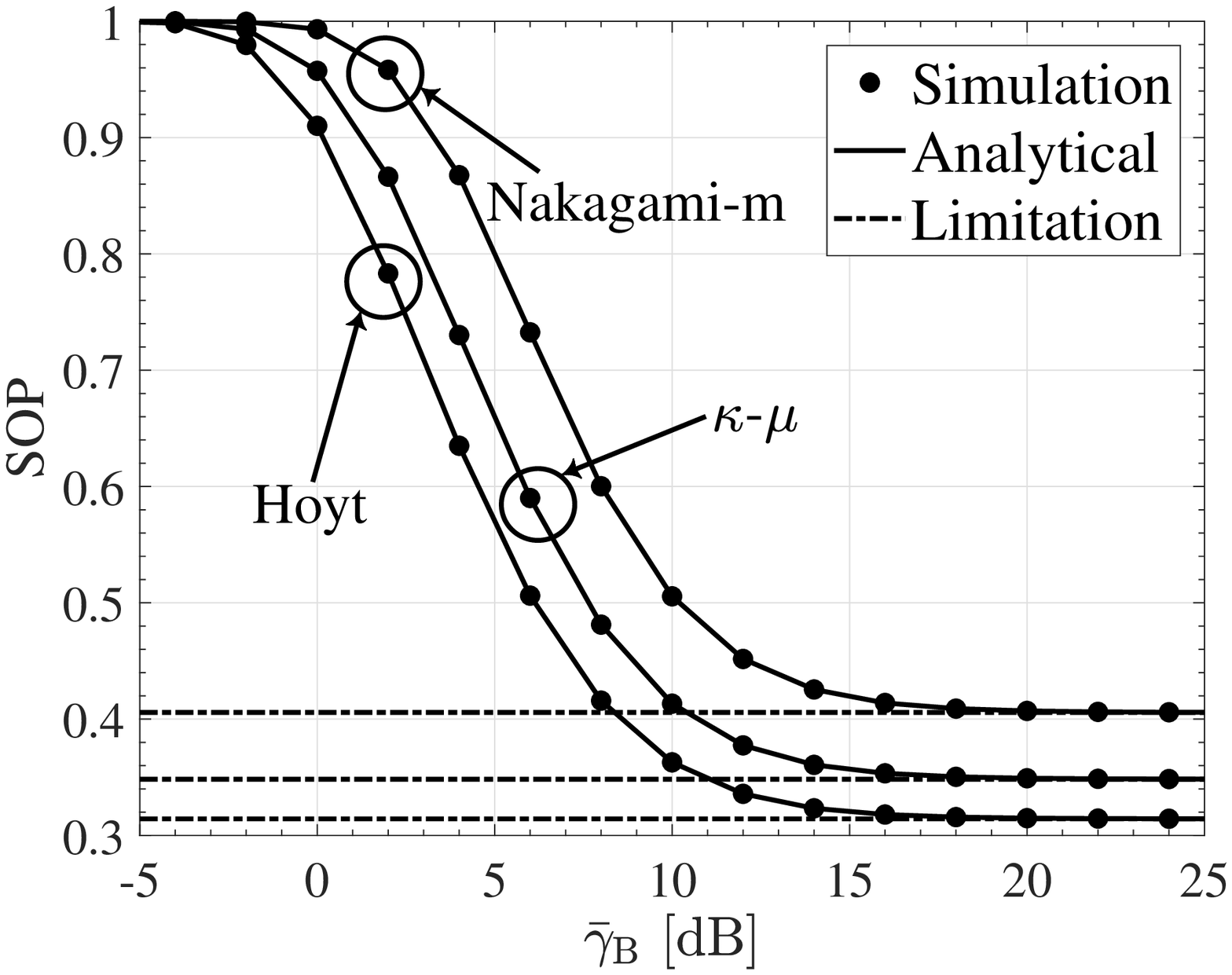}
	   \label{fig4a}	
    }
    \hspace{-14pt}
   \subfigure[$P_{\text{con}}$ versus $\bar\gamma_{\text B}$.]
    {
        \includegraphics[height=0.185\textwidth]{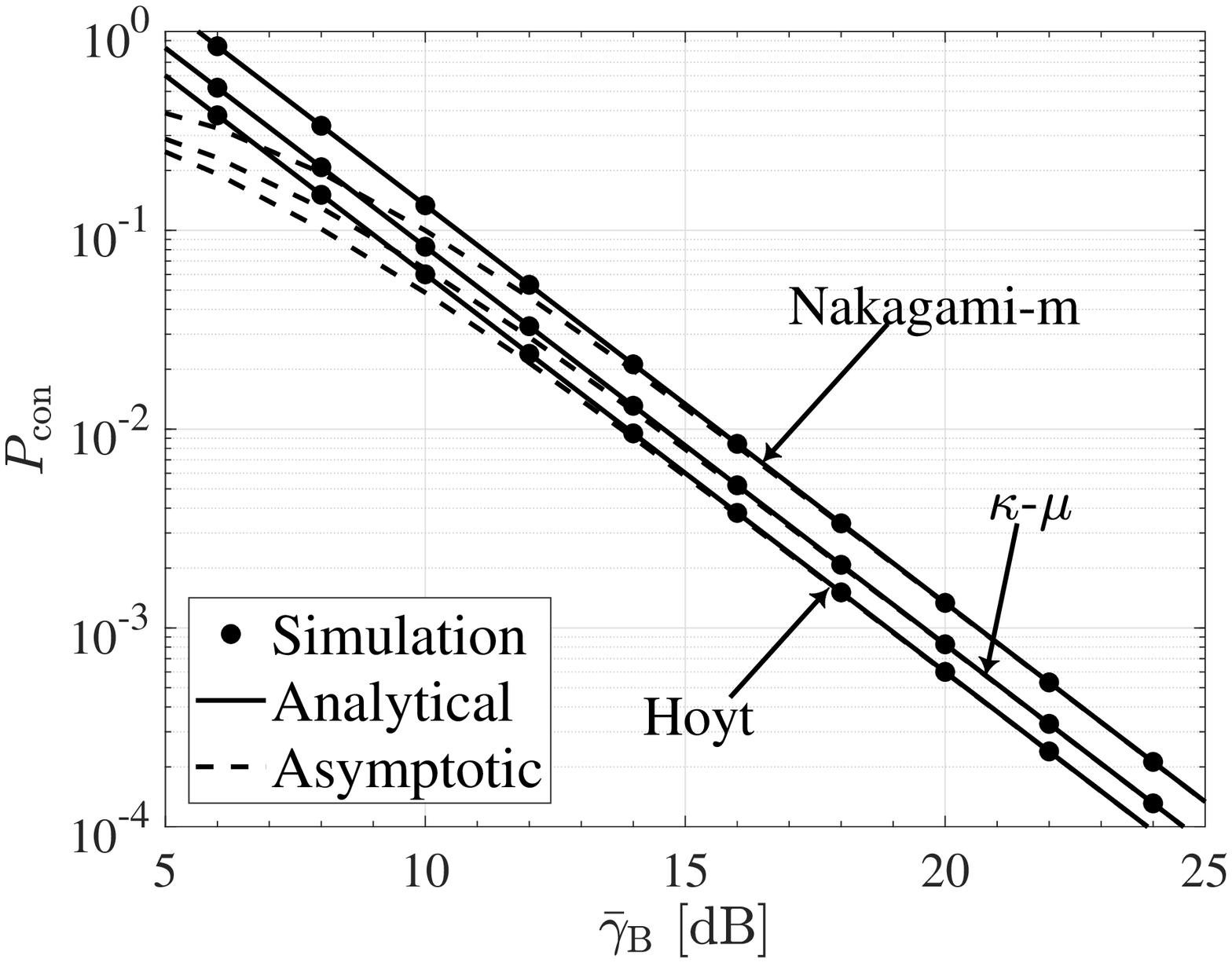}
	   \label{fig4b}	
    }
\caption{SOP of 4-QAM versus $\bar\gamma_{\text B}$ for $\bar\gamma_{\text{E}}=0$ dB and $R_{\text s}=1$ bps/Hz. The main channel undergoes $\kappa$-$\mu$ fading (${\kappa}_{\text{B}}=1$, ${\mu}_{\text{B}}=2$), while the wiretap channel respectively undergoes Nakagami-$m$ (${m}_{\text{E}}=6$), ${\mathcal K}_{G}$ (${k}_{\text{E}}=3$, ${m}_{\text{E}}=3$), and $\kappa$-$\mu$ (${\kappa}_{\text{E}}=2$, ${\mu}_{\text{E}}=1$) fadings.}
    \label{figure4}
\end{figure}

In the sequel, we provide some simulation results to verify the accuracy of \eqref{EQUATION28}. {\figurename} {\ref{fig3a}} compares the analytical and simulated SOP achieved by $M$-QAM signals over Generalized-$K$ (${\mathcal K}_G$) fading channels. The analytical SOP is calculated by \eqref{EQUATION28}, where $v$ is set as 30. As shown, the analytical results match the simulated results perfectly. For reference, the SOP achieved by Gaussian inputs is also plotted. By \cite{b4}, the SOP of Gaussian inputs converges to zero in the large limit of $\bar\gamma_{\text B}$. Yet, as discussed in Section \ref{SECTION4B}, the SOP achieved by discrete inputs converges to a positive constant, namely $1-F_{\text E}\left({\mathcal H}_M\right)$, as $\bar\gamma_{\text B}$ increases, which is consistent with the results shown in {\figurename} {\ref{fig3a}}. It can be observed from this graph that a higher modulation order yields a smaller limiting SOP, which is similar as the observation from {\figurename} {\ref{fig1a}}. Based on Section \ref{SECTION4B}, the rate of $P\left(R_{\text s}\right)$ converging to $1-F_{\text E}\left({\mathcal H}_M\right)$ equals the rate of $P_{\text{con}}$ converging to zero. To show the ROC, we plot $P_{\text{con}}$ versus $\bar\gamma_{\text B}$ in {\figurename} {\ref{fig3b}}. As shown, the derived asymptotic results accurately characterize the secrecy diversity order. Furthermore, a higher modulation order yields a slower ROC, which is consistent with the conclusion drawn from {\figurename} {\ref{fig1b}}. Actually, \eqref{EQUATION28} can be used to evaluate the SOP even though the main and eavesdropper's channels undergo different fading models. An example to verify this has been shown in {\figurename} {\ref{figure4}}, which further validates the correctness of \eqref{EQUATION28}.
\vspace{-5pt}
\section{Conclusion}
\label{section4}
{In this letter, leveraging the mathematically tractable form of the MGD model, we investigated the explicit and asymptotic secrecy performance of finite input signals over MGD fading channels. Our study provided novel insights on the wireless PLS, which may be exploited to guide future system designs.} Besides, this work established a unified and general analytical framework for evaluating the secrecy issues over wireless channels driven by finite-alphabet signals when the fading distributions can be characterized by the MGD model.
\vspace{-5pt}
\begin{appendices}
\section{Proof of Theorem \ref{theorem2}}
\label{Append1}
\vspace{-5pt}
\begin{IEEEproof}
When $R_{\text s}+{\mathcal I}_M\left(\gamma_{\text E}\right)\leq\log_2 M$, we find ${\mathcal I}_M\left(\gamma_{\text B}\right)-{\mathcal I}_M\left(\gamma_{\text E}\right)<R_{\text s}$ is equivalent to ${{\mathcal I}_M^{-1}\left(R_{\text s}+{\mathcal I}_M\left(\gamma_{\text E}\right)\right)}>\gamma_{\text B}$. Besides, when $R_{\text s}+{\mathcal I}_M\left(\gamma_{\text E}\right)>\log_2 M$, we have ${\mathcal I}_M\left(\gamma_{\text B}\right)-{\mathcal I}_M\left(\gamma_{\text E}\right)<{\mathcal I}_M\left(\gamma_{\text B}\right)-\log_2 M+R_{\text s}<R_{\text s}$, which yields ${\mathcal I}_{M}^{\text{s}}<R_{\text s}$. Taken together, we obtain $\Pr\left({\mathcal I}_{M}^{\text{s}}<R_{\text s},\gamma_{\text B}>\gamma_{\text E}\right)=P_1+P_2$, where $P_1=\int_{0}^{{\mathcal H}_M}\int_{\gamma_{\text E}}^{{\mathcal F}_M\left(\gamma_{\text E}\right)}f_{\text{B}}\left(\gamma_{\text{B}}\right)f_{\text{E}}\left(\gamma_{\text{E}}\right){\rm{d}}\gamma_{\text{B}}{\rm{d}}\gamma_{\text{E}}$, ${\mathcal H}_M={\mathcal I}_M^{-1}\left(\log_2 M-R_{\text s}\right)$, ${\mathcal F}_M\left(\gamma\right)={{\mathcal I}_M^{-1}\left(R_{\text s}+{\mathcal I}_M\left(\gamma\right)\right)}$, and $P_2=\int_{{\mathcal H}_M}^{+\infty}\int_{\gamma_{\text E}}^{+\infty}f_{\text{B}}\left(\gamma_{\text{B}}\right)f_{\text{E}}\left(\gamma_{\text{E}}\right){\rm{d}}\gamma_{\text{B}}{\rm{d}}\gamma_{\text{E}}$. In addition, $\Pr\left(\gamma_{\text B}<\gamma_{\text E}\right)$ can be written as $\Pr\left(\gamma_{\text B}<\gamma_{\text E}\right)=Z_1+Z_2$, where $Z_1=\int_{0}^{{\mathcal H}_M}\int_{0}^{\gamma_{\text E}}f_{\text{B}}\left(\gamma_{\text{B}}\right)f_{\text{E}}\left(\gamma_{\text{E}}\right){\rm{d}}\gamma_{\text{B}}{\rm{d}}\gamma_{\text{E}}$ and $Z_2=\int_{{\mathcal H}_M}^{+\infty}\int_{0}^{\gamma_{\text E}}f_{\text{B}}\left(\gamma_{\text{B}}\right)f_{\text{E}}\left(\gamma_{\text{E}}\right){\rm{d}}\gamma_{\text{B}}{\rm{d}}\gamma_{\text{E}}$. Particularly, we note that $P_2+Z_2=\int_{{\mathcal H}_M}^{+\infty}f_{\text{E}}\left(y\right)\int_{0}^{+\infty}f_{\text{B}}\left(x\right){\rm{d}}x{\rm{d}}y=1-F_{\text{E}}\left({\mathcal H}_M\right)$ and $P_1+Z_1=\int_{0}^{{\mathcal H}_M}f_{\text{E}}\left(y\right)\int_{0}^{{\mathcal F}_M\left(y\right)}f_{\text{B}}\left(x\right){\rm{d}}x{\rm{d}}y=\int_{0}^{{\mathcal H}_M}F_{\text{B}}\left({{\mathcal F}_M\left(y\right)}\right)f_{\text{E}}\left(y\right){\rm{d}}y$. As a result, \eqref{EQUATION18} can be written as $P\left(R_{\text s}\right)=P_1+P_2+Z_1+Z_2=\left(P_1+Z_1\right)+\left(P_2+Z_2\right)=1-F_{\rm{E}}\left({\mathcal H}_M\right)+\int_{0}^{{\mathcal H}_M}F_{\text{B}}\left({{\mathcal F}_M\left(y\right)}\right)f_{\text{E}}\left(y\right){\rm{d}}y$.
\end{IEEEproof}
\vspace{-7pt}
\section{Proof of Theorem \ref{theorem3}}
\label{Append2}
\vspace{-5pt}
\begin{IEEEproof}
By \cite{b19}, we have $\lim_{x\rightarrow 0^{+}}{\mathcal K}_M\left(x\right)=0$ and $\lim_{x\rightarrow+\infty}{\mathcal K}_M\left(x\right)=o\left({\rm e}^{-d_Mx}\right)$ ($d_M>0$), which together with Theorem \ref{theorem1}, yields $\left|{\mathcal M}\left[{\mathcal K}_M\left(x\right);\Lambda_{{\text B},l}+1\right]\right|<+\infty$. Besides, when $x>{\mathcal I}_{M}^{-1}\left(R_{\text s}\right)$, we have ${\mathcal W}_M\left(x\right)>0$, which implies that $\Delta_{M,l}>0$. Taken together, $\Delta_{M,l}\in\left(0,+\infty\right)$. Hence, the asymptotic SOP can be written as $P^{\infty}=1-F_{\text{E}}\left({\mathcal H}_M\right)+{G'_{a,{\text B},M}}{{\bar\gamma_{\text B}}^{-G_{d,{\text B}}}}+{o}\left({{\bar\gamma_{\text B}}^{-G_{d,{\text B}}}}\right)$, where $G'_{a,{\text B},M}=\sum_{\Psi_{{\text B},l}=\Psi_{{\text B},1}}\Delta_{M,l}\Phi_{{\text B},l}>0$ and $G_{d,{\text B}}=\Psi_{{\text B},1}>0$.
\end{IEEEproof}
\end{appendices}

\end{document}